\documentclass{aa} 
\usepackage{amssymb}
\usepackage{amsmath}
\usepackage{graphicx}
\usepackage{natbib}
\usepackage{rotating}
\usepackage{epsfig}
\usepackage{subfig}
\usepackage{colortbl}
\usepackage{lscape}
\usepackage{tabularx}
\usepackage{longtable,lscape}
\usepackage{txfonts}
\begin{document}
   \title{The parsec-scale structure of radio-loud broad absorption line quasars}


\author{G. Bruni \inst{1}
\and D. Dallacasa \inst{1,2} 
\and K.-H. Mack \inst{1}
\and F.M. Montenegro-Montes \inst{3}
\and J.I. Gonz\'alez-Serrano \inst{4}
\and J. Holt \inst{5}
\and F. Jim\'enez-Luj\'an \inst{6,7}}

   \institute{INAF-Istituto di Radioastronomia, via Piero Gobetti, 101, I-40129 Bologna, Italy\\ \email{bruni@ira.inaf.it}
   \and Università di Bologna, Dip. di Astronomia, via Ranzani, 1, I-40129 Bologna, Italy 
   \and European Southern Observatory, Alonso de C\'ordova 3107, Vitacura, Casilla 19001, Santiago, Chile
   \and Instituto de F\'isica de Cantabria (CSIC-Universidad de Cantabria), Avda. de los Castros s/n, E-39005 Santander, Spain
   \and Leiden Observatory, Leiden University, P.O. Box 9513, NL-2300 RA Leiden, The Netherlands
   \and Dpto. de F\'isica Moderna,  Universidad de Cantabria, Avda de los Castros s/n,  E-39005 Santander, Spain
   \and Centro Astron\'onico Hispano Alem\'an de Calar Alto (CAHA), C/ Jes\'us Durb\'an Rem\'on 2-2 E-04004 Almer\'ia, Spain}

   \date{}

  \abstract
   {Broad absorption line quasars (BAL QSOs) belong to a class of
     objects not well-understood as yet. Their UV spectra show BALs in
     the blue wings of the UV resonance lines, owing to ionized gas with
     outflow velocities up to 0.2 c.  
They can have radio emission that is difficult to characterize and that needs
to be studied at various wavelengths and resolutions. 
}
   {We aim to study the pc-scale properties of their synchrotron
     emission and, in particular, to determine their core
     properties.} 
   {We performed observations in the Very Long Baseline Interferometry (VLBI) technique, 
using both the European VLBI Network (EVN) at 5 GHz, and the Very Long Baseline Array (VLBA) at 5 and 8.4 GHz
    to map the pc-scale structure of the brightest
     radio-loud objects of our sample, allowing a proper morphological
     interpretation.} 
   {A variety of morphologies have been found: 9 BAL QSOs on a total
     of 11 observed sources have a resolved structure. Core-jet,
     double, and symmetric objects are present, suggesting different
     orientations. In some cases the sources can be young GPS or
     CSS. The projected linear size of the sources, also considering
     observations from our previous work for the same objects, can
     vary from tens of pc to hundreds of kpc. In some cases, a diffuse
     emission can be supposed from the missing flux-density  with
     respect to previous lower resolution observations. Finally, the magnetic
     field strength does not significantly differ from the values
     found in the literature for radio sources with similar sizes.} 
 {These results are not easily interpreted with the youth scenario
   for BAL QSOs, in which they are generally compact objects still
   expelling a dust cocoon. The variety of orientations, morphologies,
   and extensions found are presumably related to different possible
   angles for the BAL producing outflows, with respect to the jet
   axis. Moreover, the phenomenon could be present in various phases
   of the QSO evolution.} 


   \maketitle
%

\section{Introduction}
The nature and origin of broad absorption line quasars (BAL QSOs) is still an open issue in the
framework of active galactic nuclei morphology and evolution. The key characteristic of
this class of objects ($\sim$20\% of the entire QSO population, \citealt{Hewett})
resides in their UV spectra, where broad absorption lines (BALs) are
present in the blue wings of the UV resonance lines, owing to ionized
gas with outflow velocities up to 0.2 c. Some evidence of BALs in
young or luminous IR galaxies has led to proposing an evolutionary scenario
(\citealt{Briggs, Sanders}), in which this phenomenon is due to the
young age of these objects: the central AGN would still be expelling
the enveloping dust cocoon. The fraction of the QSOs showing BALs can
then be interpreted as the duration of this phase in relation to the
total active life. This view was supported by the presence of
prototypical characteristics of young radio sources, like GigaHertz Peaked Sources (GPS)
or compact steep sources (CSS), among BAL QSOs (\citealt{Montenegro, Liu}). 

An orientation model proposed by \cite{Elvis} predicts that BAL
outflows are present in all QSOs, but only when they intercept the
line of sight would the BALs be detected. In this case the percentage
of BAL QSOs would constrain the solid angle of the outflow. 
The bipolar wind model proposed by \cite{Punslya,Punslyb} was used by
\cite{Ghosh} to explain the polar BAL QSOs found by \cite{Zhou}: in this model the BAL outflow would be
aligned with the polar axis, while the relativistic jet would be nested inside of it.
Recent works, based on large samples, do not confirm a young age for all BAL QSOs, 
but also a variety of orientations have been found
(\citealt{Bruni,DiPompeo}). Also, on the arcsec-scale the majority of these sources remain
unresolved in the radio band, with only about 10\% or less showing an extended morphology
(\citealt{Bruni,DiPompeo}).

Radio-loud BAL QSOs are a small fraction of the BAL QSO population
($\sim$10\%, \citealt{Hewett,Shankar}), but the radio emission can be an additional tool for
understanding the orientation and the age (\citealt{ODea2}) of these
sources. We focus on this class of objects to collect information that
can help in understanding which of the two models is the most
probable. Thus, we embarked on a VLBI project to test the inner
structure of the QSO. 

The VLBI technique is an
important tool for this kind of study, allowing one to reach the
necessary resolution and sensitivity to study distant objects like BAL
QSOs. This approach offers different indicators of the morphology of
the radio source: (1) The detection of radio jets and their asymmetry
can provide constraints on the radio source orientation; 
(2) The
spectral index analysis allows us to discriminate between core-jet or
double structure; 
(3) Variability with respect to the total
flux-density already measured with the VLA can give an estimate of the
activity and orientation of the QSO jet axis with respect to the line
of sight. (4) In case the pc-scale structure displays mini-lobes, which are typically found in
young radio sources, the projected linear size of the radio source can
be used to estimate the kinematic age (\citealt{Dallacasa}).

In previous works on BAL QSOs, using the VLBI imaging technique (\citealt{Jiang, Kunert, Kunert2, Liu, Montenegro2,
Gawronski}),
various morphologies and sizes have been found,
and in most cases sources were unresolved even at pc-scale resolution.
In this paper we present the first results of an extensive observational
campaign to determine the pc-scale structure and morphology of a
complete sample of radio-loud BAL QSOs, whose characteristics in the radio-band 
were presented in \cite{Bruni}. 
Beyond the study of the pc-scale radio properties of the BAL QSO
population, we aim at distinguishing which of the presented
models could be the most suitable for explaining the BAL phenomenon. 

This paper is organized as follows.
In Sec. 2 we present the sources and the observations, and in Sec. 3 we
show the pc-scale radio maps  and discuss the morphology. In
Sec. 4 we discuss our results in the light of earlier works in the
literature.  

The cosmology adopted throughout the paper assumes a flat universe and
the following parameters: $H_{0}$=70 km s$^{-1}$ Mpc$^{-1}$,
$\Omega_{\Lambda}$=0.7, $\Omega_{M}$=0.3. The adopted convention for
the spectral index definition is $S_{\nu} \propto \nu^{\alpha}$. 
%
%
%
\section{Radio observations}
This work is based on a well-defined sample of radio-loud BAL QSOs
presented in \cite{Bruni}. This sample has been selected from the
fourth release of the Sloan Digitized Sky Survey (SDSS) QSO catalogue
(\citealt{Schneider07}), with a flux-density threshold of 30 mJy at
1.4 GHz and a redshift range of 1.7$<$z$<$4.7: these criteria resulted
in a sample of 25 objects.
The radio spectra of the sources, presented in \cite{Bruni}, show
that BAL QSOs are weaker at radio-frequencies above 1.4 GHz, with a few exceptions,
requiring sensitive observations to study their pc-scale structure.

In this paper, we present VLBI observations of 11 of the brighter sources 
($S_{1.4} > \sim50$ mJy) included in our sample.
These 11 are a random subset of these brighter sources.
The high-resolution study of sources 1159+01 ($S_{1.4} = 266$ mJy) and 1624+37 ($S_{1.4} = 56$ mJy), 
not included here, has already been presented by \cite{Montenegro2}: we thus obtained parsec-scale radio maps for 
a total of 13 out of 25 BAL QSOs in the whole sample. The study of the polarized flux density
was not possible in these works, since much more observing time
would have been needed to detect the low fraction of polarized flux density
of these objects.
More information on overall radio spectra and the kpc-scale
polarization properties of the objects studied here, as well as the
remaining sources in the sample, are available in \cite{Bruni}. 
In Table \ref{sources} we report some basic information on the 11 BAL
QSOs presented here, while a summary of the observations is reported 
in Table \ref{summary}.

\begin{table*}
\renewcommand\tabcolsep{3.7pt}
 \centering
  \caption{Sample of the 11 radio-loud BAL QSOs studied in this
    paper.} 
  \label{sources}
  \begin{tabular}{ccrrrcc}
  \hline
\hline
   \multicolumn{1}{c}{Name}         			&
   \multicolumn{1}{c}{SDSS name}			&
   \multicolumn{1}{c}{RA}           				& 
   \multicolumn{1}{c}{DEC}          				& 
   \multicolumn{1}{c}{z}            				&
   \multicolumn{1}{c}{$S_{1.4}$}       			&   	
   \multicolumn{1}{c}{BAL}					\\
   
   \multicolumn{1}{c}{}         				&
   \multicolumn{1}{c}{}         				&
   \multicolumn{1}{c}{(J2000)}      		& 
   \multicolumn{1}{c}{(J2000)}      		&
   \multicolumn{1}{c}{}      				&
   \multicolumn{1}{c}{(mJy/beam)} 			&
   \multicolumn{1}{c}{Type} 			\\
    
   \multicolumn{1}{c}{(1)}          		&
   \multicolumn{1}{c}{(2)}          		&
   \multicolumn{1}{c}{(3)}        		&
   \multicolumn{1}{c}{(4)}         	 	&
   \multicolumn{1}{c}{(5)}         	 	&
   \multicolumn{1}{c}{(6)}         	 	&
   \multicolumn{1}{c}{(7)}         	 	\\

\hline
0044+00     & J004444.06+001303.5 &   00 44 44.0657 			& +00 13 03.560    		&   2.28  &53.1	& -	\\
0756+37     & J075628.25+371455.6&	07 56 28.24~~~~	 	& +37 14 55.6~~~~ 	&   2.51  &239.4~~	& -	\\
0816+48     & J081618.99+482328.4&	08 16 18.9960 			& +48 23 28.469    		&   3.57 &68.3	& -	\\
0849+27     & J084914.27+275729.7&	08 49 14.2716 			& +27 57 29.709     	&   1.73 & 52.8	& Hi	\\
1014+05     & J101440.35+053712.6&	10 14 40.3561 			& +05 37 12.682     	&   2.01 & 55.0	& Hi	\\
1102+11     & J110206.66+112104.9&	11 02 06.6617 			& +11 21 04.900     	&   2.35  & 82.3	& -		\\
1237+47     & J123717.44+470807.0&	12 37 17.44~~~~ 		& +47 08 07.0~~~~ 	&   2.27  & 78.5	& FeLo	\\
1304+13	& J130448.06+130416.6&	13 04 48.0594 			& +13 04 16.590    		&   2.57  &49.6	& -		\\
1327+03      & J132703.21+031311.2&	13 27 03.21~~~~	 	& +03 13 11.2~~~~		&   2.83  &60.7	& -		\\
1406+34      & J140653.84+343337.3&	14 06 53.84~~~~	 	& +34 33 37.3~~~~		&   2.56  &164.4~~	& -		\\
1603+30      & J160354.15+300208.6&	16 03 54.1534 			& +30 02 08.705     	&   2.03  &53.7	& Hi	\\
\hline
\end{tabular} 
\begin{list}{}{}
\item[{\bf Notes:}] 
Column 3,4, and 5 are respectively RA, DEC and redshift as
    measured from the SDSS. For phase-referenced sources we give the coordinates of the brightest component peak, 
    derived from our VLBI observations.
    Column 6 gives the peak flux density at 1.4 GHz from the FIRST catalogue (\citealt{Becker, White}).
    The last column is the BAL type as given in \cite{Bruni}: a hyphen indicates that it could not be determined.
\end{list}
\end{table*}
\begin{table*}
 \centering
  \caption{Summary of the observations and instrumental setup. In the
    penultimate column the typical beam sizes are given (half-power
    beam-width).}\label{summary} 
  \begin{tabular}{clccccc}
  \hline
  \hline
  \multicolumn{1}{c}{Run}	& \multicolumn{1}{c}{Date}    & \multicolumn{1}{c}{Telescope}  & \multicolumn{1}{c}{Frequencies}	& Bandwidth & $\theta_{HPBW}$ & Time on source \\
	&	&	&	(GHz)	&	(MHz)	&	(mas)	& (hour) \\
\hline
1 & 25$-$26 Oct 2009  			& EVN 	& 5 		 & 32	& 5            & 2 \\
2 & 19, 22, 26, 28 Feb 2010 		& VLBA 	& 5, 8.4   & 64	& 4, 2        & 1, 2 \\
3 & 4, 7, 23, 26 Apr - 1, 2 May 2011& VLBA	& 5, 8.4   & 64	& 4, 2        & 3, 4 \\
\hline
\end{tabular}
\end{table*}

All the data reduction and map analysis was performed using
the NRAO AIPS\footnote{http://www.aips.nrao.edu} software, following
the usual procedures once editing of raw visibilities and of antenna temperature (TSYS)
measurements were carried out. Based on the variation in the amplitude
solutions once system temperatures and antenna gain have been
considered, the accuracy of the flux-density scale can be estimated to
be within $\pm$5\%.
The final angular resolution of the maps is in the range 2-5 mas. 

\subsection{VLBA data}
Seven objects were observed with the Very Long Baseline Array (VLBA) at 5 and 8.4 GHz 
(C and X-band, respectively) with a total bandwidth of 64 MHz at each
frequency, in two observing runs in 2010 and 2011. In the first run an average 
on-source time of about one hour at 5 GHz and two hours at 8.4 GHz was
allowed. The correlation was performed with the VLBA correlator at
the National Radio Astronomy Observatory (NRAO) in
Socorro (US). In the second run, when the fainter sources were observed, a 
total of three hours at 5 GHz and four hours at 8.4 GHz were spent on each
target. The data were then processed with the new Distributed FX (DiFX) software
correlator (\citealt{Deller}). Given that, in general, the target
BAL QSOs were weak; the VLBA observations were carried out
in phase referencing mode for most of sources; the transfer of delay, phase, and phase rate
solutions from a reference source close to target allowed absolute positions to be 
obtained, which significantly improved the earlier
information taken from the FIRST catalogue (\citealt{Becker, White}) .
Therefore, the phase centre in the maps presented in Figure
\ref{VLBI_1} refers to the shift applied and not to the peak
in the map plane whose position is instead reported in Table
\ref{sources}. 
Since the correlation was done using the position from the FIRST
survey, we had to allow a rather wide field for the initial imaging of
each target source. In fact we started with about 1$''$ wide
maps, and once the target source was detected, a smaller field was
selected after applying appropriate shifts in RA and
DEC in order to have the target source at about the centre of the
map. The same shift was applied at both frequencies. Then a few
iterations of phase self-calibration were performed. The final maps
are shown in Fig. \ref{VLBI_1} and are discussed in the next section. 

\subsection{EVN data}

Four additional sources were observed with the European VLBI Network
(EVN) at 5 GHz (C-band), in 2009, using the antennas in Jodrell Bank, Westerbork, Effelsberg,
Onsala, Medicina, Torun, Shanghai, Urumqi, Noto, Yebes, and the two
MERLIN telescopes in Darnhal and Knockin. A total bandwidth of 32
MHz and an average on-source time of about two hours were allowed. The
data were correlated at the Joint Institute for VLBI in Europe (JIVE) correlator in Dwingeloo (The Netherlands). 

For three of the sources we carried out phase-referencing.
The maps thus adopt the same convention as for the VLBA ones. 
 Standard procedures, including fringe-fitting, were used for the data
reduction of the target sources. Delay, phase, and rate solutions were
generally found above the standard signal-to-noise ratio threshold,
for all the target sources.

%
%
%
%
%
%
%
%
%
%
\section{Results}

Once final maps had been obtained, we carried out a 2-D Gaussian fitting
solving for component position, flux-density, size, and position
angle. The resulting information is reported in Table
\ref{fluxes}. The distance between the centroids of each individual
component as found by the fits were used to calculate the total
projected angular and linear sizes of the source.   

\begin{figure*}
\begin{center}
\includegraphics[width=17cm]{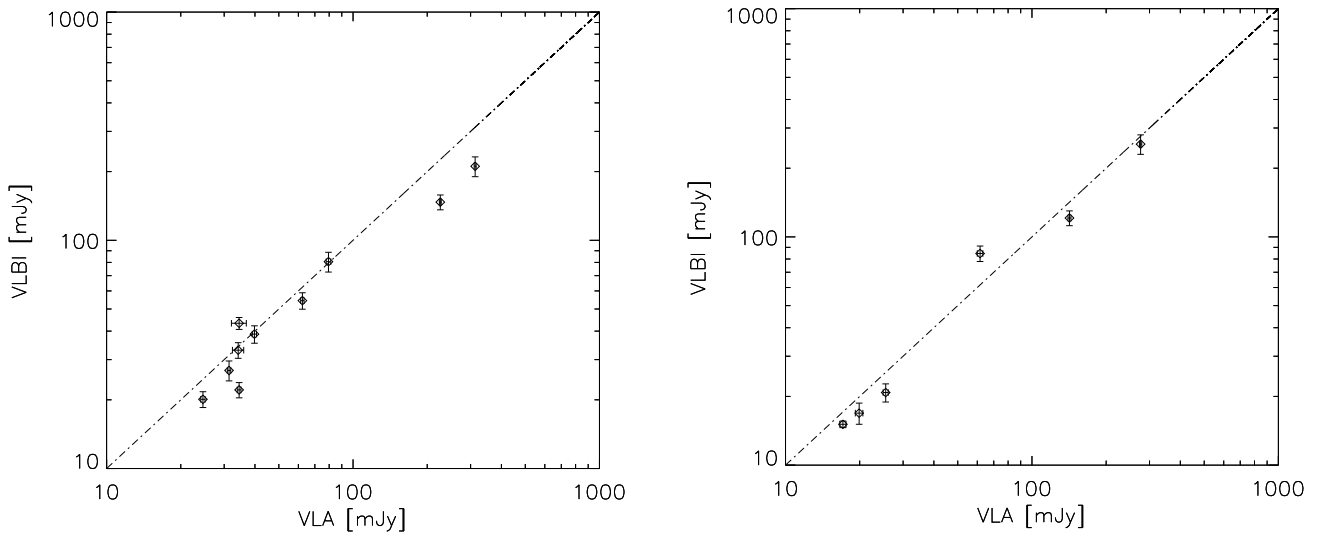}
\caption{Flux densities of VLBI vs VLA at 5 (left) and 8 GHz (right), for all of the sources with both measurements. The dashed
line indicates a slope equal to 1. The outliers ($\sigma_{var}>3$) are sources 0756+37, 1406+34 and 1014+05 at 5 GHz, and 1237+47 at 8.4 GHz.} 
\label{variability}
\end{center}
\end{figure*}

In the following, we briefly describe each source, observed with the VLBA 
(Section 3.1) or the EVN (Section 3.2). For the
often mentioned overall radio-spectrum of the sources we refer to
\cite{Bruni}. Figure \ref{variability} provides a comparison with the VLA flux-densities presented 
in our previous work. In this section we also discuss any hint of variability for the individual sources.
Finally, in Table \ref{sources} we provide the position of the brightest component peak
for the seven sources with absolute position information from phase-referencing.

\subsection{VLBA maps}

In this section, the pc-scale structure of the BAL QSOs observed
with the VLBA is briefly discussed. Most of the sources have been found
to be unresolved in the earlier arcsecond-scale observations. The typical
resolution of the VLBA maps is $\sim$~4 mas at 5 GHz and $\sim$~2 mas at 
8.4 GHz. The spectral indices $\alpha_{5}^{8.4}$ for the
individual components were calculated using the integrated flux density from the Gaussian fit, and do not refer any to local values.  
The availability of maps at two frequencies often helps in 
correctly interpreting the source morphology in terms of cores or
steep spectrum components.

\begin{figure*}
\begin{center}
\includegraphics[width=14cm]{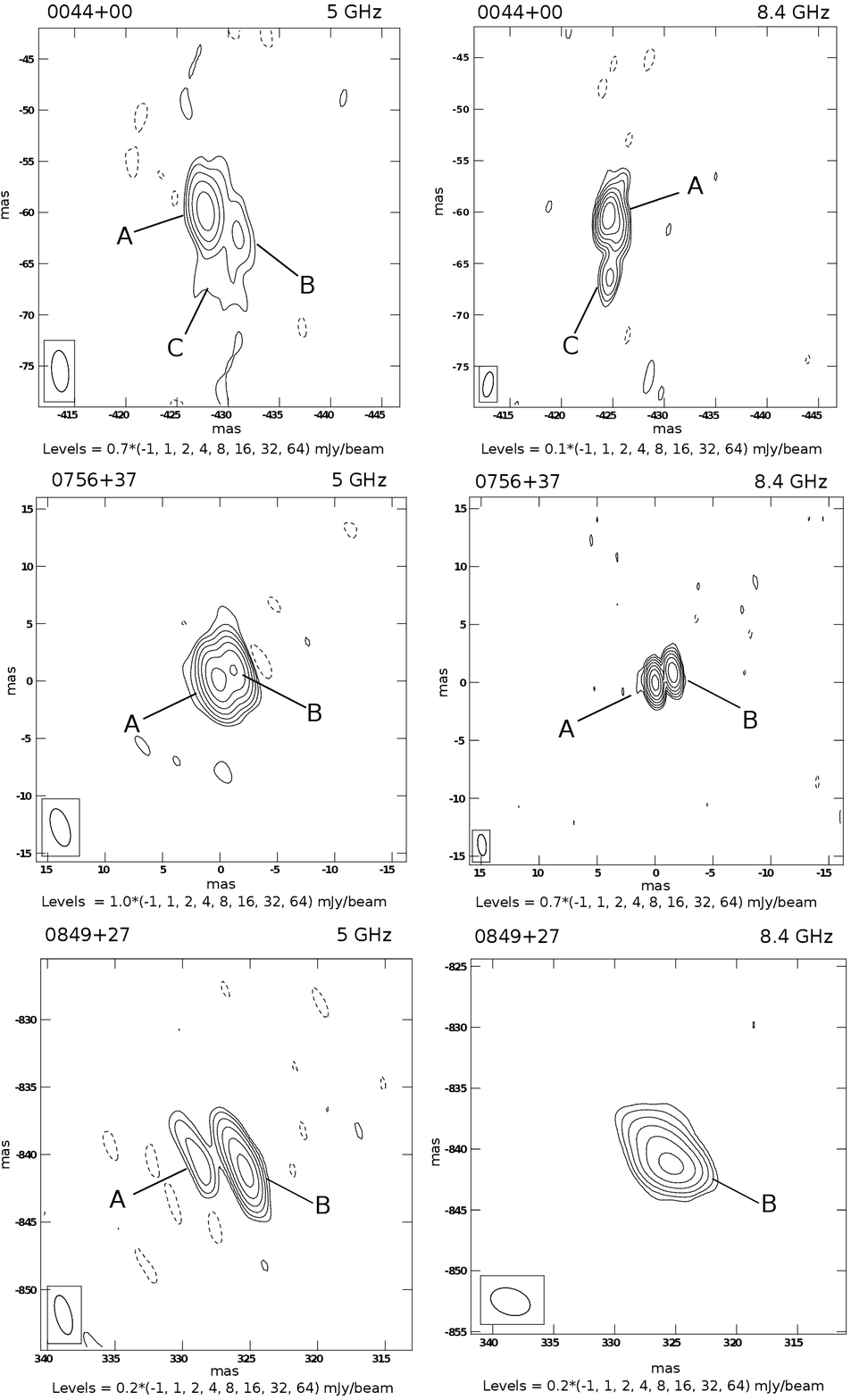}
\caption{Maps of the 6 BAL QSOs resolved with the VLBA. The synthesized beam size is shown
  in the lower left corner of the map. Levels are 3-$\sigma$
  multiples, according to the legend. Orientation is N-up, E-left.} 
\label{VLBI_1}
\end{center}
\end{figure*}

\addtocounter{figure}{-1}
\begin{figure*}
\begin{center}
	\includegraphics[width=14cm]{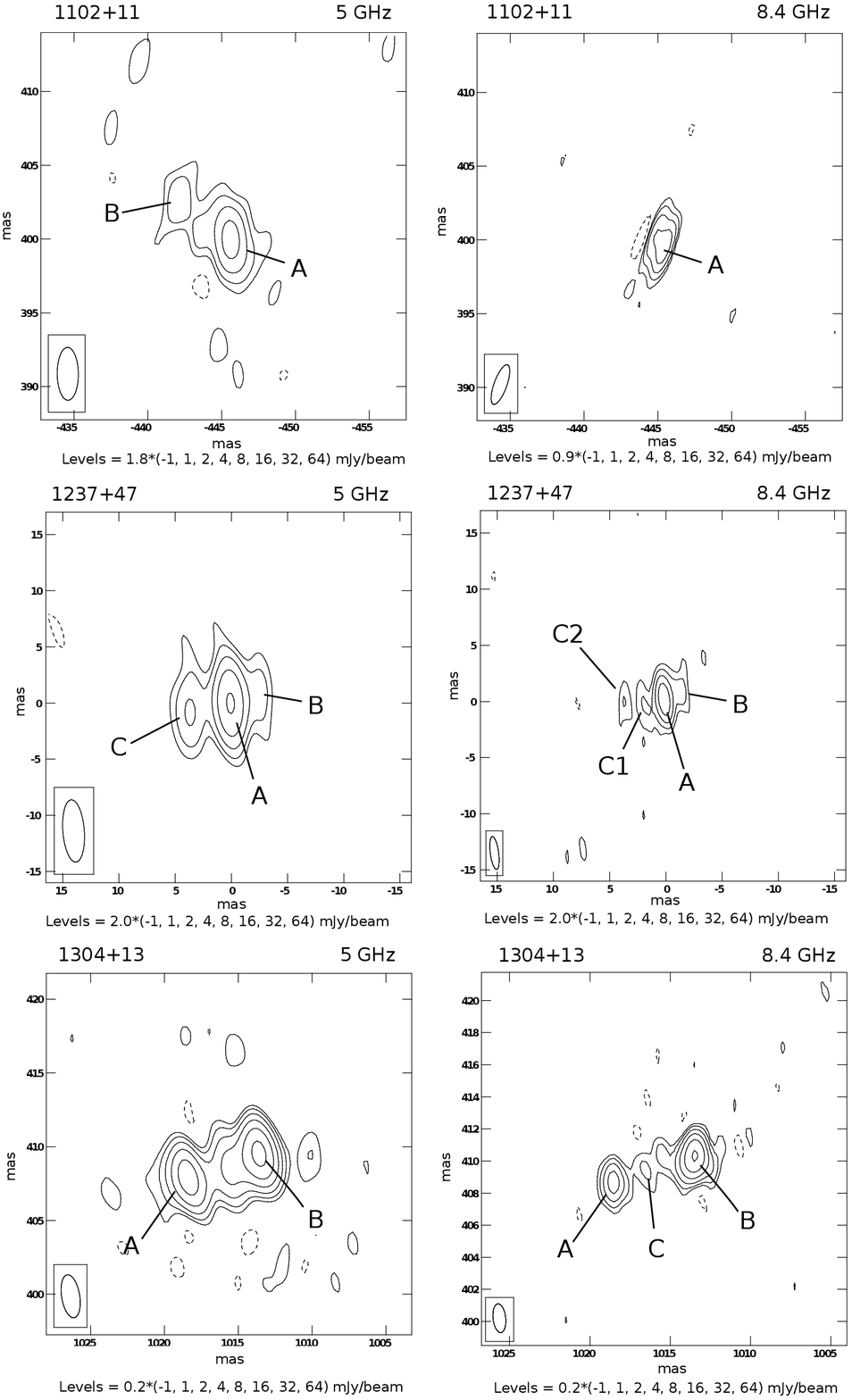}
	\caption{Continued.}
\label{VLBI_2}
\end{center}
\end{figure*}

\subsection*{0044+00}

Both maps at 5 GHz and 8.4 GHz show a resolved structure. In particular
at 5 GHz, three misaligned components are present: two of them are
clearly visible, the third is embedded in some diffuse
structure. While a Gaussian fit was possible for components A and B,
for component C it did not converge since it barely stands against some
diffuse emission, and only an upper limit for its flux-density could
be obtained. 
In case the map at 8.4 GHz was not available, component C would
 not have been considered. In this 8.4 GHz map, much deeper
than that at 5 GHz, only components A and C are visible, suggesting a
steep spectral index for component B whose surface brightness
drops below the detection level. 
Component A has a rather flat spectral index ($-$0.52$\pm$0.27)
and could be most probably classified as the core region, in which
part of the jet is also visible, as suggested by the elongation in
p.a. $\sim176^\circ$. 

Although it is not straightforward, the morphology of this source can
be interpreted as a core--jet.

\subsection*{0756+37}
In both VLBA bands we resolve the radio source into two components,
separated by $\sim$1.8 mas ($\sim$15 pc at the redshift of the object)
at the highest resolution, in p.a. $\sim-60^\circ$. 
Both components have similar sizes and spectral indexes
($-$0.38$\pm$0.27 for A and $-$0.37$\pm$0.27 for B). 
Since the overall spectrum peaks at 2.5 GHz (8.7 GHz
rest frame), and it is steep in the optically thin (power-law, not self-absorbed) part of the
emission, we can infer that we have detected two hot spots in the
mini-lobes of a typically young radio source.  

The VLBA can account for a fraction of the total flux density measured
in earlier VLA observations ($\sim$65\% at 5 GHz and  $\sim$85\% at 8.4 GHz): 
considering the variability significance as defined by \cite{Zhou}, 
the discrepancy is significant at 5 GHz, with a $\sigma_{var}$=7.0, 
while at 8 GHz we obtain $\sigma_{var}<3$. The time interval between VLA
and VLBA observations is about seven months. 
Thus, either the source is variable -- but this would be inconsistent
with the hot-spot scenario -- or, more likely, there is some
emission on angular scales not sampled by the VLBA (tens of mas or more). 

This source was observed also by \cite{Gawronski} using the EVN at 1.6 GHz, and reporting a probable
core-jet structure, but no maps were provided in that work.

\subsection*{0849+27}
This source has already been resolved in the FIRST survey (VLA at 1.4 GHz,
see \citealt{Bruni}) showing three components with a projected linear
size of 382 kpc, and a position angle for the structure of $\sim45^\circ$.  
These VLBA observations were designed to study the central component,
which was unresolved on the arcsecond scale.
The maps show two components at 5 GHz (A and B) separated by 3.5 mas
(30 pc), while only B is visible at 8.4 GHz, with a spectral index of
$-$0.12$\pm$0.20 (flat) that can be considered as the source
core. This component is resolved at both frequencies in 
p.a. $\sim30^\circ$.

\subsection*{1102+11}

The map at 5 GHz shows a structure resolved into two components,
which are separated by $\sim$4.1 mas ($\sim$34 pc), in p.a. $-53^\circ$. At 8.4
GHz only the strongest component (A) has been detected, and it turned out
to be resolved along the same direction. Its spectral index is
$-$0.97$\pm$0.29. Such a steep spectrum makes it unlikely that component
A is the source core. It is also too steep to be an active
hot spot. Component B has probably an even steeper spectral index,
since the flux density is below the 3-$\sigma$ significance limit at 8.4 GHz. 
All this makes the morphological classification of the pc-scale
structure rather unclear.\\  
The source presents a convex radio spectrum, although more
observations at low frequencies are needed to clearly determine the
turnover frequency, if present.

\subsection*{1237+47}
This source is resolved into three components at 5 GHz, with the
strongest one in the centre (A). 
At 8.4 GHz, the eastern component C is resolved in two subcomponents (C1
and C2), both aligned with A and B. Component A has an inverted
spectral index (0.41$\pm$0.29) and thus could be interpreted as the
core, while B and C1 present elongated emission toward A, with a similar
flux density at 8.4 GHz. It could be the inner part of the jets, with
an additional hot spot (C2) visible in the direction of C1. The
projected linear size from X band is 4.8 mas (40 pc).\\ 
From our previous radio observations of this source we know that it
presents a flat spectrum, which can possibly be approximated with a
convex shape with a peak at about 3.5 GHz.  
Thus, this object can be a young radio source in which the core region
is still the brightest in the source. Moreover, at 8.4 GHz its flux density 
exceeds ($\sim$37\%) the total flux density in our earlier VLA
observation. This could indicate that there is still a very active
phase in which new components are likely to be formed on short
timescales, creating the variability in the flux density. The time
interval between VLA and VLBA observations is about seven months, with a
$\sigma_{var}$=3.4, just above the 3-$\sigma$ significance threshold.\\

\subsection*{1304+13}
It was possible to resolve this source both at 5 GHz and at 8.4 GHz: in
the maps, the two brightest components are on the edges of the radio
source (A and B), while the central component C is distinguishable only at 8.4 GHz.
The fit did not converge for this last component, so we could only estimate the
flux density upper limit.

The total projected linear size of the source at 8.4 GHz is 5.3 mas
(44 pc). The overall spectrum is steep, and does not present a peak in
the GHz range, which is the only one that could be investigated at the moment.

\subsection*{1406+34}

This source is unresolved in both bands with the VLBA. We can set an
upper limit of 1.8 mas (14 pc) to its size. The map at 8.4 GHz shows a noise pattern 
that is not related to any extended structure.
The pc-scale emission
revealed by the present observations has an inverted spectrum
($\alpha$=0.36$\pm$0.26), and we are therefore seeing the optically thick,
self-absorbed part of the synchrotron emission.

The overall radio spectrum peaks at about 5.4 GHz. The VLBA flux-density
measurements at 5 GHz accounts only for $\sim$67\%  that of the VLA ($\sigma_{var}$=4.8), owing either to some intrinsic
variability or to an additional component fully resolved on the
pc-scale. Such a component, with a steep spectrum, may be responsible
for the 327 MHz emission detected in the WENSS survey
(\citealt{Bruyn}), which could be the remnant of an earlier episode of
radio activity and which also modifies the spectral slope at low
frequencies. The source is also unresolved in VLA maps.

\begin{figure}
\begin{center}
\includegraphics[width=7cm]{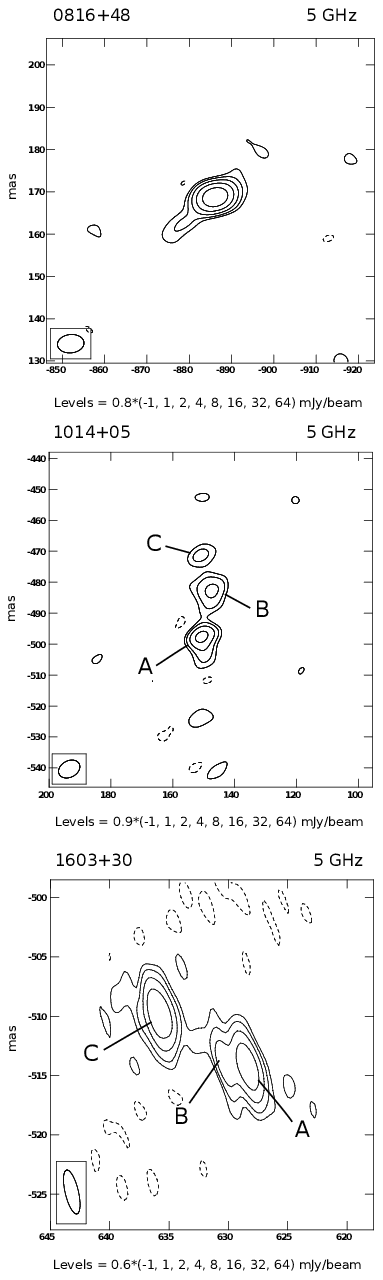}
\caption{Maps of the 3 BAL QSOs resolved with the EVN. The synthesized beam size is shown in
  the lower left corner of the map. Levels are 3-$\sigma$ multiples,
  according to the legend. Orientation is N-up, E-left.} 
\label{VLBI_1}
\end{center}
\end{figure}


\subsection{Results from EVN observations}

The EVN observed four BAL QSOs at 5 GHz (C-band). This array is made of
various radio telescopes with substantially different performances. In
particular, the presence of a large dish like the Effelsberg-100m
allows better baseline sensitivity than on a typical VLBA baseline.
However, due to the geographical distribution of the telescopes, the uv-plane is not uniformly sampled, and, on average,
the structural information that can be derived here is poorer than the
maps presented in the previous section. The typical resolution of
these maps is $\sim$5 mas.  

\subsection*{0816+48}

This source shows an elongation toward SE. 
From the major axis of the Gaussian fit (6.8 mas) we can infer an
upper limit of projected linear size of 51 pc.  
A clear classification of the spectrum was not possible in our previous work, 
but the VLA map at 1.4 GHz already showed an
elongation toward SW, not corresponding with the one present here, and
with a major and minor axes of 29 and 14 arcsec, corresponding to 217
and 105 kpc, respectively.  
This could be another case of intermittent activity with a change in
the jet orientation between the two phases. Further observations at
the intermediate VLA resolution could confirm this hypothesis.  
\subsection*{1014+05}
This source presents a complex morphology whose proper classification
is rather difficult, because
the brightest component (A) can be interpreted as the core region, and
there are two more components (B and C) toward north. 
The overall structure is well aligned in p.a. $\sim$0$^\circ$.

The distance between the centroids of A and C components is $\sim$26.6
mas, corresponding to a projected linear size of 229 pc. In our
earlier work we found a power-law spectrum for this source, but more
data below 1.4 GHz are needed to exclude any turnover in the MHz
range.\\  
Our EVN map can account for $\sim$59\% of the total flux density
measured in the VLA map ($\sigma_{var}$=8.2), since this discrepancy may arise either from
some flux-density variability or from an additional structure not sampled by
the EVN spacings.

\subsection*{1327+03}
This is an unresolved source. 
The upper limit for the linear size, estimated from
the Gaussian fit, is 46 pc (5.9 mas) and 11 pc (1.4 mas), respectively,
on the major and minor axes, but the beam is considerably elliptical in
this map so this can only be an indicative value. A power-law spectrum
down to 1.4 GHz does not show any turnover for this source. 
\subsection*{1603+30}
In this case a rather symmetric structure can be seen on the parsec
scale.
The south-western component can be fitted with two (A and B)
Gaussians. The separation between A and C (at the north-eastern edge)
components is 8.6 mas, with a  projected linear size of 74
pc.  \\ 
This source presents a convex radio spectrum, with a peak frequency at
1.4 GHz, and thus could be classified as a young GPS source, in which
the emission is dominated by the hot spots.\\
  
On the arcsecond scale, a VLA map at 22 GHz shows a resolved
structure in which some additional emission is present at about 2 arcsec
(17 kpc) to the south.
Such a linear size is more typical of CSS sources (1-20 kpc),
although the overall spectrum is typical of GPS sources. The structure
seen with the EVN is not aligned with the one seen with the VLA: 
probably a merger or a jet precession could justify the different morphologies of
the same object on different scales. This is another case of a possibly young source, like 1237+47, with a flux density at 
VLBI resolution exceeding the earlier VLA observations at the same frequency ($\sim$25\%), 
but in this case the variability significance $\sigma_{var}$ is only 2.5, i.e. below the threshold. 
This source has already been observed with EVN+MERLIN at 1.6 GHz (\citealt{Liu}), 
but owing to the lower resolution (13.0 $\times$ 4.0 mas) it was unresolved.

\begin{landscape}
\begin{table}
\renewcommand\tabcolsep{3.5pt}
  \caption{Results for the 11 radio-loud BAL QSOs studied in this paper.}  
  \label{fluxes}
  \scalebox{0.92}{
  \begin{tabular}{cccccccccccccccccc}
  \hline
  \hline
   \multicolumn{1}{c}{Name}         			&
   \multicolumn{1}{c}{comp.}           			&
   \multicolumn{1}{c}{$S_{\rm5}^{VLBI}$} 		&  
   \multicolumn{1}{c}{$S_{\rm8.4}^{VLBI}$} 	& 
   \multicolumn{1}{c}{$\alpha$}           			&
   \multicolumn{1}{c}{$S_{\rm4.8}^{VLA}$} 	& 
   \multicolumn{1}{c}{$S_{\rm8.4}^{VLA}$} 	& 
   \multicolumn{1}{c}{Maj. axis}           		&
   \multicolumn{1}{c}{Min. axis}           		&
   \multicolumn{1}{c}{Maj. axis}           		&
   \multicolumn{1}{c}{Min. axis}           		&
   \multicolumn{1}{c}{P.A.}           				&
   \multicolumn{1}{c}{LS}           		&
   \multicolumn{1}{c}{LS}           		&
   \multicolumn{1}{c}{Morphology}         		&
   \multicolumn{1}{c}{$u_{min}$}	&
   \multicolumn{1}{c}{$B_{eq}$}		&
   \multicolumn{1}{c}{$B'_{eq}$}		\\

   \multicolumn{1}{c}{}          			&
   \multicolumn{1}{c}{}          			&
   \multicolumn{1}{c}{(mJy)}        		&
   \multicolumn{1}{c}{(mJy)}         	&
   \multicolumn{1}{c}{}     			&     
   \multicolumn{1}{c}{(mJy)}          	&
   \multicolumn{1}{c}{(mJy)} 		&
   \multicolumn{1}{c}{(mas)} 		&
   \multicolumn{1}{c}{(mas)}          		&
   \multicolumn{1}{c}{(pc)} 		&
   \multicolumn{1}{c}{(pc)}          		&
   \multicolumn{1}{c}{(degrees)}        	&
   \multicolumn{1}{c}{(mas)}          	&
   \multicolumn{1}{c}{(pc)}          		&
   \multicolumn{1}{c}{}          			&
   \multicolumn{1}{c}{(10$^{-6}$erg/cm$^{3}$)} 	&
   \multicolumn{1}{c}{(mG)} 			&
   \multicolumn{1}{c}{(mG)} 			\\

   \multicolumn{1}{c}{(1)}          		&
   \multicolumn{1}{c}{(2)}          		&
   \multicolumn{1}{c}{(3)}        		&
   \multicolumn{1}{c}{(4)}         	 	&
   \multicolumn{1}{c}{(5)}     			&     
   \multicolumn{1}{c}{(6)}          		&
   \multicolumn{1}{c}{(7)} 			&
   \multicolumn{1}{c}{(8)}          		&
   \multicolumn{1}{c}{(9)} 			&
   \multicolumn{1}{c}{(10)}          		&
   \multicolumn{1}{c}{(11)}           		&
   \multicolumn{1}{c}{(12)} 			&
   \multicolumn{1}{c}{(13)}          		&
   \multicolumn{1}{c}{(14)}           		&
   \multicolumn{1}{c}{(15)}          		&
   \multicolumn{1}{c}{(16)}          		&
   \multicolumn{1}{c}{(17)}          		&
   \multicolumn{1}{c}{(18)}          		\\

\hline
0044+00   	&   	&	33.0$\pm$2.6	&	20.8$\pm$1.9	&-						&	 34.2$\pm$1.8	& 25.5$\pm$0.6	&-  	&-  	&-	&-							&			&5.6		&47	&CJ	& 				\\		
	  		& A &	25.0$\pm$2.5	&	19.0$\pm$1.9	&$-$0.52$\pm$0.27	&		    			&					&1.3&0.7&11&~~6						&	176		&			&	&	&   1.1		&	3.4	&	2.4				\\
	  		& B 	&	~~6.5$\pm$0.8	&	-				&-						&		    			&					&4.5&1.5&37&13						&	~~~~8	&			&	&	&   0.4		&	2.0	&	1.5			\\
	  		& C &	$<$2.4			&	~~1.8$\pm$0.2	&$>-$0.5				&		    			&					&2.9&0.7&24&~~6						&	171		&			&	&	&   1.1		&	3.4	&	2.4				\\
0756+37   	&   	&	147$\pm$11	&	121$\pm$9~~	&-					&	 226$\pm$2~~ 	& 142$\pm$2~~		&-	&-	&-	&-							&			&1.8		& 15&D & 				\\		
	  		& A &	88.2$\pm$8.8	&	72.2$\pm$7.2	&$-$0.38$\pm$0.27	&		    			&					&0.3&0.1&~~2&~~1						&	~~~~6	& 			&	&	&	7.7		&	9.1	&	5.7			\\
	  		& B 	&	59.2$\pm$5.9	&	48.8$\pm$4.8	&$-$0.37$\pm$0.27	&		    			&					&0.6&0.5&~~5&~~3						&	~~~~8	& 			&	&	&	2.6		&    5.3	&	3.5			\\
0816+48   	&   	&	26.9$\pm$2.7	&	-				&-						&	 31.4$\pm$0.5 	& 19.5$\pm$0.5	&3.3&0.1&24&~~1						& 108		&6.8		&51&CJ	& 	10.5~~	&    10.6~~ &	6.5			\\		
0849+27   	&   	&	14.8$\pm$1.3	&	11.8$\pm$0.2	&- 						&	-   				& -					&-	&-	&-	&-							&			&3.5		&30	& CJ& 				\\		
	  		& A &	~~2.2$\pm$0.3	&	$<$0.6			&$<-$2.5				&		    			&					&$<$2.9&$<$1.1&$<$25&$<$9		&	~~26	&			&	&	&	$>$0.4		&	$>$2.0	&	$>$1.5			\\
	  		& B 	&	12.6$\pm$1.3	&	11.8$\pm$0.2	&$-$0.12$\pm$0.20	&	    				&					&2.7&0.4&23&~~3						&	~~48	&			&	&	&	1.4		&    3.9	&	2.7			\\
1014+05   	&   	&	20.4$\pm$1.6	&	-				&-						&34.5$\pm$0.6		& 25.1$\pm$0.5 	&-	&-	&-	&-							&			&26.6~~	& 229~~	&M	& 		\\		
	  		& A &	10.3$\pm$1.1	&	-				&-						&		    			&					&$<$7.3&$<$5.3&$<$61&$<$45		& 155		&			&		&	&	$>$0.1		&	$>$1.0	&	$>$0.8		\\
	  		& B 	&	~~7.4$\pm$1.0	&	-				&-						&	    				&					&6.7&2.6&56&22						& 165		&			&		&	&	0.2		&	1.4	&	1.1		\\
	  		& C &	~~2.7$\pm$0.6	&	-				&-						&	    				&					&2.9&0.9&24&~~8						& 122		&			&		&	&	0.5		&	2.3	&	1.7		\\
1102+11   	&   	&	38.8$\pm$3.4	&	16.9$\pm$1.8	&-						&39.8$\pm$0.8		& 19.9$\pm$0.7	&-	&-	&-	&-							&			&4.1		& 34	&D	& 			\\		
	  		& A &	28.1$\pm$3.0	&	16.9$\pm$1.8	&$-$0.97$\pm$0.29	&		    			&					&$<$2.8&$<$0.9&$<$23&$<$7		& 167		&			&	&	&	$>$1.0		&	$>$3.3	&	$>$2.3			\\
	  		& B 	&	10.7$\pm$1.7	&	$<$5.4			&$<-1.3$				&	    				&					&2.4&1.2&20&10						& 10		&			&	&	&	0.5		&	2.4	&	1.7			\\
1237+47   	&   	&	54.4$\pm$4.5	&	84.6$\pm$6.6	&-						&62.3$\pm$1.0		& 61.6$\pm$1.6	&-	 &-	&-	 &-							&			&4.8		& 40&M	& 				\\		
	  		& A &	37.8$\pm$4.0	&	46.9$\pm$5.0	&0.41$\pm$0.29		&		    			&					&1.2&0.8&10&~~7						&	~~~~5	&			&	&	&				\\
	  		& B 	&	11.0$\pm$1.5	&	16.1$\pm$2.6	&0.72$\pm$0.40		&	    				&					&1.8&0.6&15&~~5						&	178		&			&	&	&				\\
	  		& C &	11.4$\pm$1.7	&	-				&-						&	    				&					&-	 &-	&-	 &-							&	-		&			&	&	&				\\
	  		& C1&	-				&	14.3$\pm$2.9	&-						&	    				&					&2.6&1.6&21&13						&	~~11	&			&	&	&				\\
	  		& C2&	-				&	~~7.3$\pm$1.7	&-						&	    				&					&2.2&0.3&18&~~2						&	179		&			&	&	&				\\
1304+13	&   	&	20.1$\pm$1.6	&	15.1$\pm$0.5	&-						&24.6$\pm$0.5		& 17.1$\pm$0.5	&-	&-	&-	&-							&			&5.3 		& 44&CJ	& 				\\		
	  		& A &	~~7.2$\pm$0.7	&	~~3.2$\pm$0.3	&$-$1.54$\pm$0.26	&		    			&					&0.8&0.5&~~6&~~4						&	~~~~3	&			&	&	&	2.0		&	4.6	&	3.2			\\
	  		& B 	&	12.9$\pm$1.4	&	11.2$\pm$0.3	&$-$0.27$\pm$0.21	&	    				&					&1.1&0.3&~~9&~~2						&	174		&			&	&	&	4.0		&	6.6	&	4.3			\\
	  		& C & 	-				&	$<$0.9			&-						&	    				&					&-	&-	&-	&-							&	-		&			&	&	&				\\
1327+03   	&   	&	80.6$\pm$8.0	&	-				&-						&79.5$\pm$1.7		& 56.5$\pm$0.7	&$<$5.9&$<$1.4	&$<$46&$<$11	&	~~~~9	&$<$5.9	& $<$46&U	&$>$0.6		&	$>$2.6	&	$>$1.9			\\		 
1406+34   	&   	&	~211$\pm$21	&	255$\pm$25	&0.36$\pm$0.26		&313$\pm$3~~		& 276$\pm$3~~	&$<$1.8&$<$0.7	&$<$14&$<$6		&	~~~~5	&$<$1.8	& $<$14&U	& 			\\		
1603+30   	&   	&	43.3$\pm$2.6	&	-				&-						&34.5$\pm$2.4		& 26.9$\pm$0.6	&-	 &-	&-	 &-							&			&8.6		& 74&M	& 				\\		
	  		& A &	18.1$\pm$1.8	&	-				&-						&	    				&					&2.4&1.3&20&11						&	~~18	&			&	&	&	0.4		&	2.0	&	1.5			\\	
	  		& B 	&	~~8.8$\pm$1.0	&	-				&-						&	    				&					&2.4&1.4&20&12						&	~~18	&			&	&	&	0.3		&	1.9	&	1.5			\\
	  		& C &	16.4$\pm$1.6	&	-				&-						&	    				&					&2.2&1.5&18&13						&	~~15	&			&	&	&	0.3		&	1.8	&	1.4			\\
\hline
\end{tabular}}
\begin{list}{}{}
\item[{\bf Notes:}] 
In
    column 2 the component is specified, columns 3 and 4 are flux densities 
    measured with the VLBA or EVN; column 5 is the spectral
    index for the components; columns 6 and 7 are the VLA flux densities 
as presented in \cite{Bruni}; columns 8-11 are the deconvolved major
    and minor axes of the Gaussian fit, at the highest resolution map available,
    for each component. Column 12, 13, and 14 are the position angle of the
    single components (from the same map used to calculate the component size) 
    and the total projected linear size of the
    source. Column 15 is the morphology classification (CJ: core-jet;
    D: double; M: multiple; U: unresolved). Columns 16, 17, and 18 are the derived quantities
from the magnetic field estimation (see formulae \ref{govoni}, \ref{govoniB}, and \ref{Brunetti}).
\end{list}
\end{table}
\end{landscape}
%
%
%
%
%

\subsection{Magnetic field estimation}

We applied the \cite{Govoni} formulae to calculate the magnetic field intensity or strength B.
Supposing the synchrotron mechanism to be at the origin of the radio emission, 
this estimate assumes that the magnetic-field energy density $U_{B}$ and the
relativistic particles energy density $U_{part}$, contained in the emitting region,
contribute with approximately equal values to the total energy density $U_{tot}$.
This \emph{equipartition} assumption allows minimizing $U_{tot}$, providing a conservative estimate of B. The formulae for minimum energy density, $u_{min}$,
and the corresponding equipartition value for the magnetic field strength, $B_{eq}$, are the following:
\begin{multline} 
	u_{min}[\frac{erg}{cm^{3}}]  = \xi(\alpha ,\nu_{1},\nu_{2})(1+k)^{4/7}(\nu_{0[MHz]})^{4\alpha /7}(1+z)^{(12+4\alpha)/7}\\
							   \times(I_{0[\frac{mJy}{arcsec^{2}}]})^{4/7}(d[kpc])^{-4/7} 
\label{govoni}
\end{multline}
\begin{equation}
\label{govoniB}
B_{eq}=(\frac{24\pi}{7}u_{min})^{1/2},
\end{equation}
where  $z$ is the source
redshift, $I_{0}$ the source brightness at the frequency $\nu_{0}$,
$d$ the source depth, k the proton-to-electron energy ratio,
$\nu_0$ the observed frequency, and $\alpha$ the spectral index
of the synchrotron emission (adopting the convention $S_{\nu}\propto\nu^{-\alpha}$). The constant
$\xi(\alpha,\nu_{1},\nu_{2})$, depending on $\alpha$ and on the
minimum and maximum energy of the charged particles (corresponding to
$\nu_1$ and $\nu_2$), is tabulated in Table 1 of \cite{Govoni}. We chose
a value of $2.5 \times 10^{-12}$ corresponding to $\alpha=0.5$, and
minimum and maximum emission frequencies of 10 MHz and 100 GHz,
respectively. In this calculation we assumed $k$=1, a magnetic field
filling-factor $\phi$=1, and an ellipsoidal volume for the component,
with depth $d$ equal to the minor axis.

Assuming a low-energy cut-off in the particle energy distribution, the
derived magnetic field $B'_{eq}$ is related to $B_{eq}$ by the
following formula (\citealt{Brunetti}): 
\begin{equation}
\label{Brunetti}
B'_{eq}\sim\gamma_{min}^{\frac{1-2\alpha}{3+\alpha}}\cdot B_{eq}^{\frac{7}{2(3+\alpha)}}.
\end{equation}
We assumed a minimum Lorentz factor $\gamma_{min}=100$ in this calculation.

In Table \ref{fluxes} we present the results of the magnetic field
estimate for each component (or for the all the sources in case it is
unresolved).  
For the unresolved sources or components, having
deconvolved dimensions equal to zero, we used the major and minor axes
of the beam and the peak flux density. Components without Gaussian fit
were omitted. For two sources (1237+47 and 1406+34), the estimation was not
possible, since components show an inverted spectrum, which means we are looking 
at the self-absorbed part of the spectrum: in this case, the measured flux density is biased
by the absorption, and does not reflect the real emission of the component. 

We found values of $u_{min}$ ranging from 10$^{-7}$ to 10$^{-5}$
erg/cm$^3$, corresponding to $B'_{eq}$ between 0.8 and 6.5 mG. In
Sec. 4 we compare these results with values from literature. It is
worth noting that  some \emph{a priori} assumptions have been made in
this estimation. The value of $k$, the filling factor of the magnetic
field $\phi$, and the depth of the source along the line of sight $d$
can have slightly different values from the ones used here, changing the results up to an order of magnitude.


\section{Discussion}

Radio-loud BAL QSOs were considered extremely rare sources before the
advent of large surveys like the SDSS. After the release of SDSS DR4, it
has been possible to select larger samples of BAL QSOs with a
considerable emission also in the radio domain. The sample we studied
here has been defined to have the largest possible number of sources with radio flux
density high enough to allow a detailed study of the pc-scale
structure. 

In the literature, 32 BAL QSOs have to date been observed with the
VLBI imaging technique (\citealt{Jiang, Kunert, Kunert2, Liu, Montenegro2,
Gawronski}). With this work we add a further nine previously unobserved sources, improving the
statistics concerning the pc-scale radio morphology. 
Only 8 BAL QSOs out of 32 ($\sim$25\%) were found to be unresolved in
earlier works, while the remaining objects show a variety of
morphologies, including core-jet, quite symmetric, or complex
structures. In some cases, a re-orientation or jet-precession scenario
has been proposed (\citealt{Kunert, Kunert2, Gawronski}) to explain the
peculiar morphology. \cite{Doi} have observed another 22 BAL QSOs with the
Optically ConnecTed Array for VLBI Exploration (OCTAVE), subarray of the Japanese VLBI Network, 
to detect pc-scale radio emission due to non-thermal
jets. Their work does not provide maps, hence no morphological
information comparable to the previous, but also in this case the
results did not point to a single pole-on geometry for BAL QSOs. 

\begin{figure}
\begin{center}
\includegraphics[width=9cm]{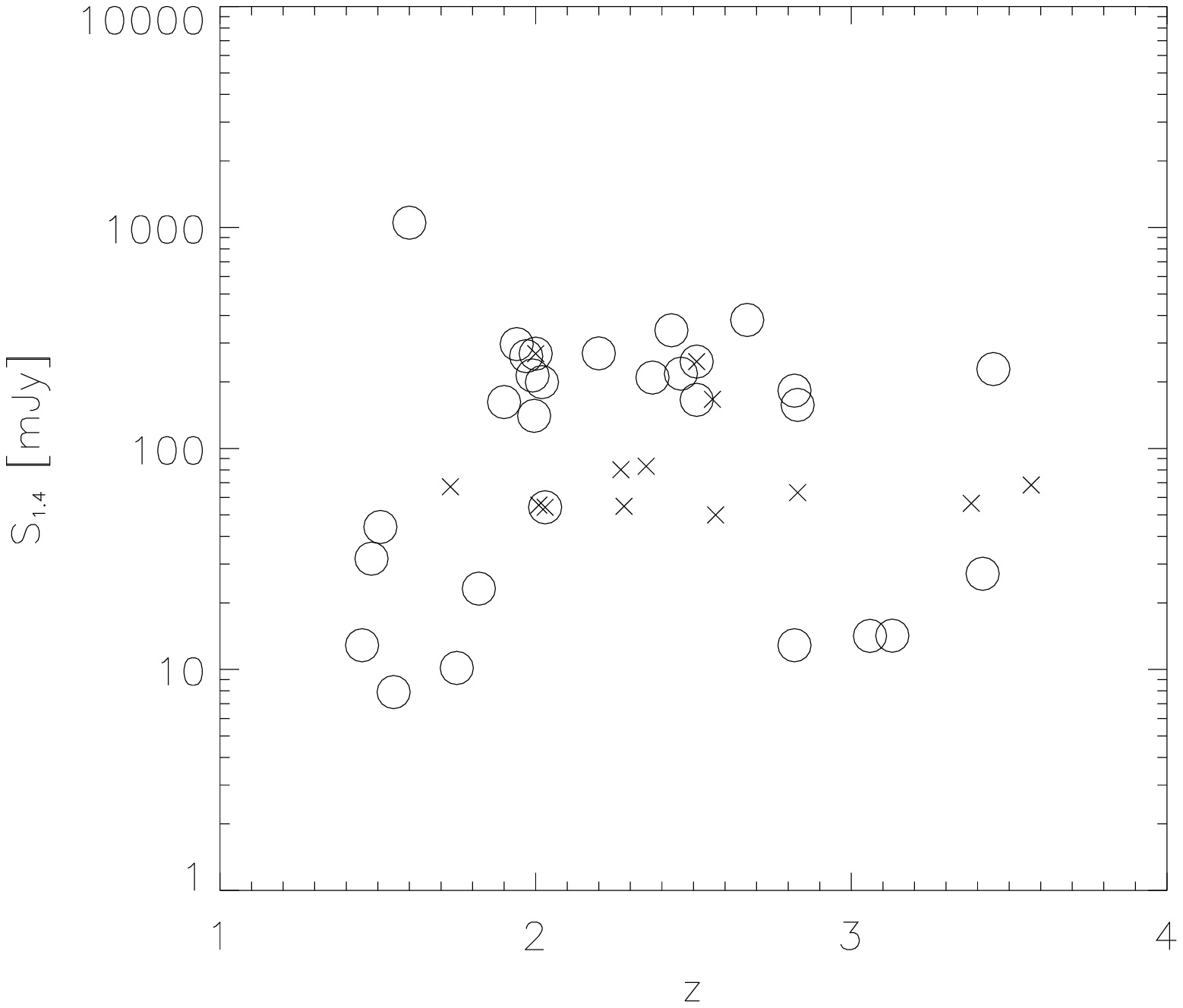}
\caption{Integrated flux density from the FIRST versus redshift for the 13 BAL QSOs with VLBI observations comprised in the \cite{Bruni} sample (from this paper or from \citealt{Montenegro2} - crosses) and from the literature (circles). For source 0849+27, resolved at the FIRST resolution, we considered the flux density given by the sum of the single components.} 
\label{samples}
\end{center}
\end{figure}

Concerning our sample, four sources (16\%) out of the 25 in the whole sample 
(0816+48, 0849+27, 1103+11, and 1603+30) were already resolved at the
intermediate angular resolution of our previous VLA observations
(\citealt{Bruni}), resulting in linear sizes of 217, 380, 69, and 17 kpc, respectively. 
The remaining 84\% of the sample turned out to be unresolved, with a linear size 
below 20 kpc in 90-95\% of the cases.
A previous VLBI study by our group (\citealt{Montenegro2}), based on a
fainter sample of radio-loud BAL QSOs presented in \cite{Montenegro},
shows that on the parsec scale a variety of morphologies are found,
implying  different orientations. 
In that work a total of five sources were observed with the VLBA,
resulting in two resolved sources (40\%). In this work we present
pc-scale observations of eleven sources: nine of them present a
resolved structure ($\sim$82\%) without a predominant morphology. Both
double sources with a radio spectrum consistent with being young (and
implying jets oriented to large angles to the line of sight) and
core-jet objects (with their axis quite aligned to the line of sight)
have been found, reinforcing the idea that orientation is not the
dominant factor in determining the properties of BAL QSOs.
In our pc-scale images, three sources (0756+37, 1014+05, and 1406+34)
can account only for part of the VLA flux density measured at the same frequency. 
This missing flux density could be related to some emission
with angular scales not sampled by the VLBI baselines (i.e. exceeding
several tens of milliarcseconds, and therefore resolved out by the
observations presented here) but contributing to the arcsecond scale
measurement. Otherwise, such a difference may arise from intrinsic
variability of the most compact components. All of these three sources were unresolved in the VLA maps,
with linear size below 8 kpc.

If we also consider the sources 1159+01 and 1624+37, studied by
\citealt{Montenegro2},
we end up with a total of 11 out of 13 ($\sim$85\%) sources that
appear resolved on the pc-scale. In 1159+01
a central core was found, with two faint symmetric extensions and
another two components toward SW, while in the case of 1624+37 a
core-jet structure was detected. This is consistent with the variety
of morphologies we have found in the present work, and it suggests a
variety of orientation with respect to the line of sight, in agreement
with the results in \cite{Bruni}. 

These results, as a whole, agree with the findings
available in the literature, both in terms of unresolved sources
fraction and in possible morphologies (and orientations).  
Figure \ref{samples} shows the integrated flux density from the FIRST
versus redshift for the 13 objects with VLBI observations in the
sample from \cite{Bruni} and from the literature. 
Both redshift and flux density ranges are comparable to the literature, 
so we can safely include the results from this paper in the context of
previous VLBI observations of BAL QSOs. 

After a decade of VLBI observations, BAL QSOs do not seem to be
predominantly unresolved, or to present any preferred morphology. All
of them have been detected on the pc-scale. In some cases these
sources can be compact, with a projected linear size below 20 kpc, or
even extremely compact, $<$1 kpc, but there are examples of extended
sources (hundreds of kpc), like 0849+27, presenting BAL features, or
cases of BAL QSOs with traces of a diffuse emission, probably related
to a low-frequency component remnant of an earlier active phase. Thus,
the interpretation of BAL QSOs as young sources seems to be
overly simplistic. On the other hand, the variety of jet axis
orientations with respect to the line of sight we found does not allow
us to explain this particular subset of QSOs as likely to be seen
from a particular line of sight, as suggested by the orientation model
(\citealt{Elvis}). The outflows at the origin of the BAL features
should thus be present at various orientations with respect to the
jet axis, and in various phases of the QSO evolutionary track.   

We determined the magnetic field intensity in the various emitting
regions, under the conventional assumptions of pure synchrotron
emission and minimum total energy (i.e. the equipartition field), and
we found values of a few mG. \cite{Dallacasa2,
  Dallacasa3} present VLBA, EVN and Merlin 
observations of a sample of compact steep-spectrum sources (CSS) drawn
from the B3-VLA sample (\citealt{Vigotti}) and covering about the same
radio luminosity range of the BAL QSOs presented here. They calculated
the minimum energy density and the magnetic field strength, adopting
the same assumptions and methodology, for radio components with sizes
and flux densities comparable to the ones presented in this
work. They found $u_{min}$ in the range between 10$^{-8}$ and 10$^{-7}$
erg/cm$^3$, corresponding to $B'_{eq}<3$ mG, comparable with
our findings.  
This suggests that the radio-emitting plasma in radio-loud BAL QSOs
has physical conditions similar to other (non-boosted) sources, while
typical values for large-scale radio sources, such as FRI and II, spans from tens to hundreds of $\mu$G (\citealt{Croston, Stawarz1, Stawarz2}).


\section{Conclusions}
We have presented the results of EVN and VLBA observations of 11 among
the brightest radio-loud BAL QSOs in our sample. A pc-scale imaging,
together with a spectral index analysis of the components, was
realized. We can summarize our conclusions as follows.
\begin{itemize}

\item Nine out of eleven sources (82\%) present a resolved structure,
  and various morphologies are visible: double, core-jet and symmetric
  structures have been found, and therefore different orientations can
  be inferred. The percentage of resolved sources is comparable to
  the general results from earlier works in the literature.\\ 

\item The projected linear sizes of the sources presented here range
  from a few pc (upper limit for the unresolved sources) to several
  tens to several hundred pc. A few sources have additional radio emission
  on much larger scales, as is visible in the VLA observations, and
  therefore a typical radio size of BAL QSOs does not exist. These objects explore 
the same parameter space as the common radio sources (radio galaxies and radio quasars).
In the original sample from \cite{Bruni}, $\sim$80\% of sources have linear sizes below 20 kpc,
still indicating that, on the arcsec-scale, the majority of BAL QSOs remain unresolved.

The missing flux density between VLA and VLBI observations can
  suggest in some cases a diffuse emission that is too faint to be detected
  with these observations. In one case (1406+34) this can be due to a
  low-frequency component, probably the remnant of an earlier radio
  activity of the AGN. 

  A complex history can also be supposed for
  other sources, when a re-oriented jet seems to be at the origin of
  the complex morphology found.\\   

\item The variety of linear sizes and spectral characteristics found
  in these VLBI observations and from our previous work
  (\citealt{Bruni}) seems to exclude a simple explanation for all BAL
  QSOs as young compact objects. At the same time, given the variety
  of morphologies that can be found, a particular orientation is
  unlikely to be present: the outflows triggering BAL features
  are then likely to be present in both young and older QSOs, at
  various orientations with respect to the jet axis.\\ 

\item A calculation of the minimum energy density and magnetic field
  strength shows similar values to the ones present in the literature,
  suggesting that no particular conditions are present in the radio-emitting regions of
  BAL QSOs. 

\end{itemize} 
%
%
%


\begin{acknowledgements}
We would like to thank M. Mahmud for kindly helping us with the
preparation of the EVN schedules.\\ 
Part of this work was supported by a grant of the Italian Programme
for Research of Relevant National Interest (PRIN No. 18/2007, PI:
K.-H. Mack).\\ 
The authors acknowledge financial support from the Spanish Ministerio
de Ciencia e Innovaci\'on under project  AYA2008-06311-C02-02 and AYA2011-29517-C03-02.\\
The European VLBI Network is a joint facility of European, Chinese,
and other radio astronomy institutes funded by their national research
councils.\\ 
The National Radio Astronomy Observatory is a facility of the National
Science Foundation operated under cooperative agreement by Associated
Universities, Inc. \\
This work made use of the Swinburne University of Technology software
correlator, developed as part of the Australian Major National
Research Facilities Programme and operated under licence.\\  
\end{acknowledgements}


\end{document}